\documentclass{article}

\usepackage{arxiv}
\usepackage{natbib}
\usepackage[utf8]{inputenc} 
\usepackage[T1]{fontenc}    
\usepackage{hyperref}       
\usepackage{url}            
\usepackage{booktabs}       
\usepackage{amsfonts}       
\usepackage{nicefrac}       
\usepackage{microtype}      
\usepackage{lipsum}
\usepackage{graphicx}
\usepackage[]{babel}
\usepackage{threeparttablex}
\usepackage{longtable}
\graphicspath{ {./images/} }

\title{Trust and Trustworthiness from Human-Centered Perspective in HRI – A Systematic Literature Review}

\author{
 Debora de Souza \\
School of Digital Technologies\\
  Tallinn University\\
  \texttt{deboracs@tlu.ee} \\
   \And
 Sonia Sousa \\
School of Digital Technologies\\ Tallinn University \& Taltech\\
  \texttt{scs@tlu.ee} \\
  \And
 Kadri Kristjuhan-Ling\\
School of Engineering \\
Tallinn University of Technology (Taltech) \\
\texttt{kadri.kristjuhan-ling@taltech.ee}\\
  \And
Olga Dunajeva \\
Virumaa College  \\
Taltech \\
 \texttt{olga.dunajeva@taltech.ee} \\
   \And
Mare Roosileht \\
Virumaa College  \\
Taltech \\
 \texttt{mare.roosileht@taltech.ee} \\
   \And
Avar Pentel \\
Virumaa College  \\
Taltech \\
 \texttt{avar.pentel@taltech.ee} \\
   \And
Mati Mottus \\
Virumaa College  \\
Taltech \\
 \texttt{ mati.mottus@taltech.ee} \\
   \And
Mustafa Can Ozdemir\\
Virumaa College  \\
Taltech \\
 \texttt{mustafa.ozdemir@taltech.ee} \\
   \And
Zanna Gratsjova\\
Virumaa College\\
Taltech\\
\texttt{zanna.gratsjova@taltech.ee} \\
}

\begin{document}

\maketitle

\begin{abstract}
The transition from Industry 4.0 to Industry 5.0 highlights recent European efforts to design intelligent devices, systems, and automation that can work alongside human intelligence and enhance human capabilities. Human-machine interaction in this vision goes beyond simply deploying machines, such as autonomous robots, for economic advantage; it also demands a well-defined human-machine interaction (HMI). Such understanding of how Industry 5.0 visions align with user preferences and needs to feel safe while collaborating with autonomous intelligent systems takes priority. It requires not just a societal and educational shift in how we perceive technological advancements but also demands a human-centric research vision. Aligned with this perspective, we conducted a systematic literature review focusing on how trust and trustworthiness can be characteristics of humans and systems in Human-Robot Interaction (HRI) as we move towards the affluent Industry 5.0. This review aims to provide an overview of the most common methodologies and measurements and collect insights about barriers and facilitators for fostering trustworthy HRI. After a rigorous quality assessment following the Systematic Reviews and Meta-Analyses guidelines, using rigorous inclusion criteria and screening by at least two reviewers, 34 articles were included in the review. The findings underscores the significance of trust and safety as foundational elements for promoting secure and trustworthy human-machine cooperation. Confirm that almost 30\% of the revised articles do not present a definition of trust, which can be problematic as this lack of conceptual clarity can undermine research efforts in addressing this problem from a central perspective. It highlights that the choice of domain and area of application should influence the choice of methods and approaches to fostering trust in HRI, as those choices can significantly affect user preferences, as well as their perceptions and assessment of robot capabilities. Additionally, this lack of conceptual clarity can act as a potential barrier to fostering trust in HRI and explains the sometimes contradictory findings or choice of methods and instruments used to investigate trust in robots and other autonomous systems in the literature. 

\end{abstract}


\section{Introduction}
Our project is a competition on Kaggle (Predict Future Sales). We are provided with daily historical sales data (including each products’ sale date, block, shop price and amount). And we will use it to forecast the total amount of each product sold next month. Because of the list of shops and products slightly changes every month. We need to create a robust model that can handle such situations.

Industry 5.0 relies on deploying intelligent automation systems that prioritize human-machine interactions (HMI) and optimize data-driven tools and processes. Such a vision aligns with the recent European efforts to design intelligent devices, systems, and automation that complement human capabilities \cite{Dixson,Pinto2022} 
It also shares a common goal: promote a societal change in which advanced technologies, such as autonomous robots, "are actively used in everyday life, industry, healthcare and other spheres of activity, not primarily for economic advantage but for the benefit and convenience of each citizen" \cite{nahavandi2019industry}. In sum, the objectives envisioned by Industry 5.0 contemplate that technologies adapt to human needs and the diversity of human nature, empower workers, and improve their process efficiency rather than replace them 
\cite{nahavandi2019industry}.
Enabling tools to augment the human ability to be creative and solve problems, benefiting from machine precision and data processing while preserving the human role in critical decision-making processes.  
By that token, current Human-Robot Interaction (HRI) approaches should focus on human and system qualities that affect the potential of building robust and adaptive industrial ecosystems to safeguard this human-centric vision of technology development. Drawing on the HRI-related studies, we examined the phenomena of trust and the factors that enable and hinder trust in intelligent automation in contexts where humans and machines coexist in meaningful human-machine collaboration. 
The following sections of this work describe the literature review's systematic methodological approach, including the research objectives and strategies. Then, illustrate how the article identification, selection, and data extraction process was performed and validated, followed by an analysis and synthesis of the definitions of trust employed, the origins and focus of the studies, and the aspects listed as facilitators and barriers. Finally, discuss a pathway to support the successful implementation of autonomous technologies and foster the successful adoption of technology from a human-centered design perspective (HCD).

\subsection{Research scope}
Reflecting on the above-described human-centric transition, one of the essential challenges is designing autonomous robotic technologies that promote seamless human-machine collaboration. Current definitions of autonomous robotics describe it as autonomous systems capable of independently determining their actions without human intervention \cite{gebru2022trustreview, abeywickrama2023specifying}. 

However, with the current fast technological development pace, the range and capabilities of such technological innovations vary from using robots in factories, at home, driverless transports, and autonomous drones to deploying COBOTS as teammates, checking orders with robot concierges, etc. A strict separation and role distinction between humans and machines may not be sufficient to mitigate the risks of unintended consequences when aiming to foster effective human-machine collaboration in complex and high-risk application contexts. 

Therefore, promoting ethical, socially responsible, and user-trustworthy automation may require alternative design approaches. Adopting a Human-centric approach to the problem is referred to as a solution for leveraging an effective, safe, and trustworthy human-robot collaboration \cite{europeancommission2021regulation, Dixson, sousa2024human}.

\subsubsection{Industry 5.0 vision}
While Industry 5.0 envisions human-centric and sustainable innovations, the rapid adoption of robotics and AI raises new challenges. For example, such fast development in technological innovations introduces concerns about trust, ethics, safety, and security when human-machine collaboration is required \cite{naiseh2022trustworthy, abeywickrama2023specifying}. Although we are aware that the principle of jidoka has been around for a hundred years and, when combined with Just-in-Time (JIT) principles, forms a robust foundation for lean manufacturing, ensuring efficiency, flexibility, and quality in production processes. Thus, providing an automated system with human supervision indicates a) poor machine maintenance and b) poor work organization \cite{krijnen2007toyota}. 
With current technological advances, the concept of combining automated systems with human supervision, often referred to as \emph{"Human in the Loop"} (HITL) automation, is widely discussed in automation and robotics. This approach emphasizes that human oversight can enhance the efficiency and reliability of automated systems. It is argued that this combination leads to better machine maintenance and improved work organization. HITL automation ensures that humans can intervene when necessary, providing critical judgment and decision-making that machines alone might not handle effectively. Such a discussion of the extent to which machines can have human judgment is widely debated and depends much on the actions and context in which the machine works. For simple and routine tasks, following the principle of jidoka and separating human and machine work can improve safety and product quality and reduce time.
Regardless, the \emph{"Human in the Loop"} framework requires new design approaches and methods to ensure that humans supervise, guide, and intervene in system operations when needed. This ensures that critical decisions are aligned with ethical, contextual, and situational requirements, and humans are ethically accountable for their choices, such as AI-generated clinic diagnoses. It can also foster a partnership between humans and machines, such as when drivers might take over autonomous vehicles and control them during complex traffic scenarios. This ensures system safety and reliability, particularly in high-risk applications. 

\subsubsection{Human-centric vision}
Human-Computer Interaction (HCI) researchers have been adopting the human-centered approach to the problem since 1970, with the first focus on improving system usability to make interactions between humans and machines more efficient and effective. The field remains interdisciplinary, drawing from computer science, psychology, design, and other areas to improve how people interact with technology \cite{rogers2014diffusion}. Additionally, it contemplates studies on the emotional and social aspects, following the principles of cognitive psychology to provide a deeper understanding of technology-related human behavior. Today, it continues to advance in the field and plays a crucial role in related subfields such as Human-Robot Interaction (HRI), Human-Machine Interaction (HMI), and Human-Agent Interaction (HAI).  

HCI's interdisciplinary methodologies facilitate the design and evaluation of technologies from a human-centric perspective, adhering to guidelines like ISO 9241-210:2019. These frameworks promote usability, usefulness, and user experience, making HCI vital to achieving Industry 5.0's sustainability goals, inclusivity, and human-centric innovation. 

\subsubsection{Human-centric trust vision}

In this interplay, trust is essential for enabling the successful integration, adoption, and cooperation of humans with autonomous robotic technologies to achieve better outcomes. As these technologies develop, they become increasingly capable of performing complex cognitive tasks and taking over more activities previously handled by humans. Their acceptance becomes more difficult as people's feelings of losing control grow, along with the idea that such innovations necessitate new knowledge, skills, and an understanding of their operations \cite{brynjolfsson2017artificial} The \emph{"black box"} nature of AI systems further complicates this relationship, fueling debates about their benefits and risks \cite{aihleg2019ethics, gulati2019hctm, sousa2023challenges}. 
Moreover, research indicates that ensuring both physical safety (e.g., through robust sensor-based systems) and psychological safety (e.g., by fostering user trust and well-being) is crucial for the successful integration of robotic systems \cite{gihleb2022industrial,abeywickrama2023specifying, Interact_Analysis, towers2019keep, laux2024trustworthy}. Furthermore, the concept of trust in human-robot interaction remains complex and often misapplied, leading to confusion about its role and definition. This lack of clarity has impeded research to foster acceptance and trust in autonomous technologies such as self-driving cars, drones, and collaborative robots \cite{visser2020overtrust, PILACINSKI2023e18164}.

In sum, to realize Industry 5.0's full potential, it is essential to address these gaps by developing systems that prioritize human-centric design and a deep understanding of psychological, social, and ethical dimensions. This will enable effective human-machine partnerships, fostering a future where technological progress aligns with societal and environmental well-being.  
\section{Methodology}



\begin{figure}[!ht]
\includegraphics[width = 12cm ]{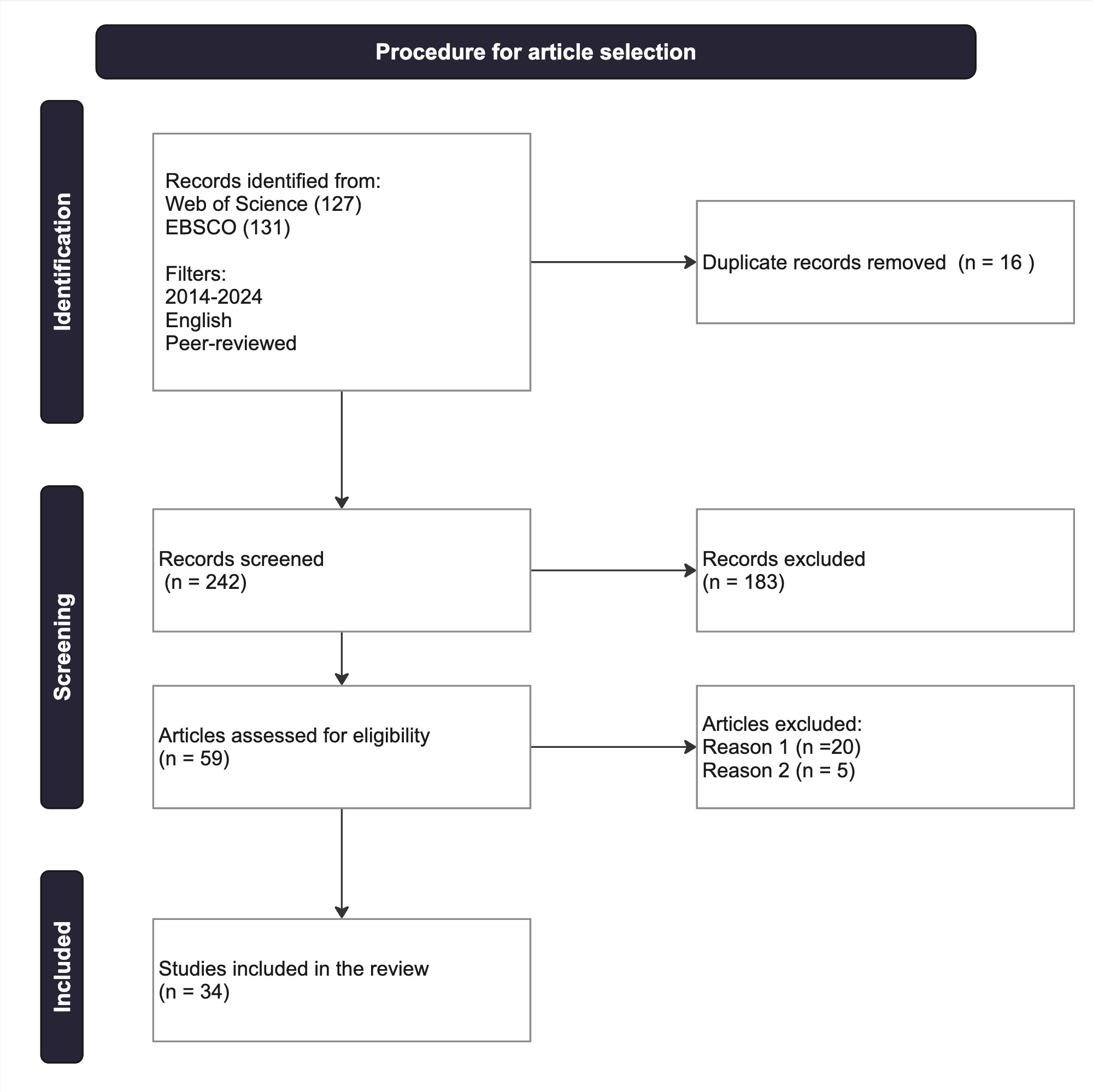}
\caption{The PRISMA flow chart of the study selection process.\label{fig1}}
\end{figure}

This systematic literature review followed the guidelines preferred reporting items for Systematic Reviews and Meta-Analyses guidelines (PRISMA) for qualitative synthesis \cite{moher2009preferred} and structured the process as in the following steps: a) establishing the research objectives and defining search string terms, inclusion, and exclusion criteria; b) identification from research databases; c) screening titles and abstracts; d) reviewing selected articles for data extraction;  and e) analysis, categorization and summary of results. The flowchart presented in Figure~\ref{fig1}  illustrates the stages of the screening process following \cite{moher2009preferred}. 

The literature review main goal is to explore how trust is assessed and conceptualized within varied studies from human-robot interaction literature. In parallel, this study reflects on the methods employed to evaluate trust in HRI systems, commenting on the focus, user groups, and applications researched. This study aims to answer three main questions:

\quad RQ1: What are the most common methodologies to study users’ trust in HRI?

\quad RQ2: What has been the focus of HRI researchers when investigating trust?

\quad RQ3: What are the barriers and facilitators for fostering trustworthy HRI?

The procedure presented in the following subsections delineates the stages of the study. 

\subsection{Literature Search and Search Strategy}

Pursuing that goal, the search strategy was carried out on the same day (March 27, 2024) and captured and was guided by similar reviews exploring human-machine trust aspects \cite{9720720, bach2022reviewtrustai, Pinto2022}.
The query included terms such as \emph{"trust assessment"}, \emph{"trustworthiness"}, and \emph{"user trust"}. Additionally, the search was expanded to include literature on \emph{"human-centered computing"}, \emph{"user-centered design"}, as well as \emph{"HCI assessments methods"}. Terms within similar categories were combined with OR, and then the results from each category were mixed with AND. 

This plural focus was used to integrate perspectives on user trust from both technological and human-centered design lenses. 

The articles were searched via Tallinn University databases using the libraries of Web of Science (127) and EBSCO (131), with constraints for publication year (i.e., 2014-2024), peer-reviewed articles, and English. 

\unskip

\subsection{Screening}

The records identified were organized into a poll of 258 articles. After duplicates were removed (N=16), the articles were perused by title and abstract following the criteria established for inclusion and exclusion presented in Figure~\ref{fig2} below. Disagreements were resolved through third review members and discussions. 
The resulting list contained the first and second authors’ checked 59 articles eligible for inclusion. As previously illustrated in Figure~\ref{fig1}, 25 other articles were removed from the review sequentially due to a) Reason 1: the lack of systematic description of the approach used; and b) Reason 2: the application examined did not refer to robots but to personal virtual assistants or recommendation system.

We evaluated the quality of the selected articles using a mixed methods tool, then extracted data to meet the review’s objectives. The data was combined and summarized in two ways: as numerical summaries for the quantitative data and as a narrative for the qualitative data.

\subsection{Data extraction}

A set of aspects were enlisted as relevant information that should be observed in the article review, piloted by the authors, and later iterated. The goal was to ensure a standard procedure and simplify the data extraction process. The data extracted included the following information: author(s) reference, the study aims, trust definition, research focus, study type, methods applied, trust metrics, and target behaviors. Reviewers were also invited to take notes of pertinent information that could benefit answering the research questions. Additionally, a second round of data extraction was carried out to collect further study-specific data, which was meant to inform the thematic framework built by the first and last authors within the analysis phase.  

\begin{figure}[!ht]
\includegraphics[width = 11.5cm ]{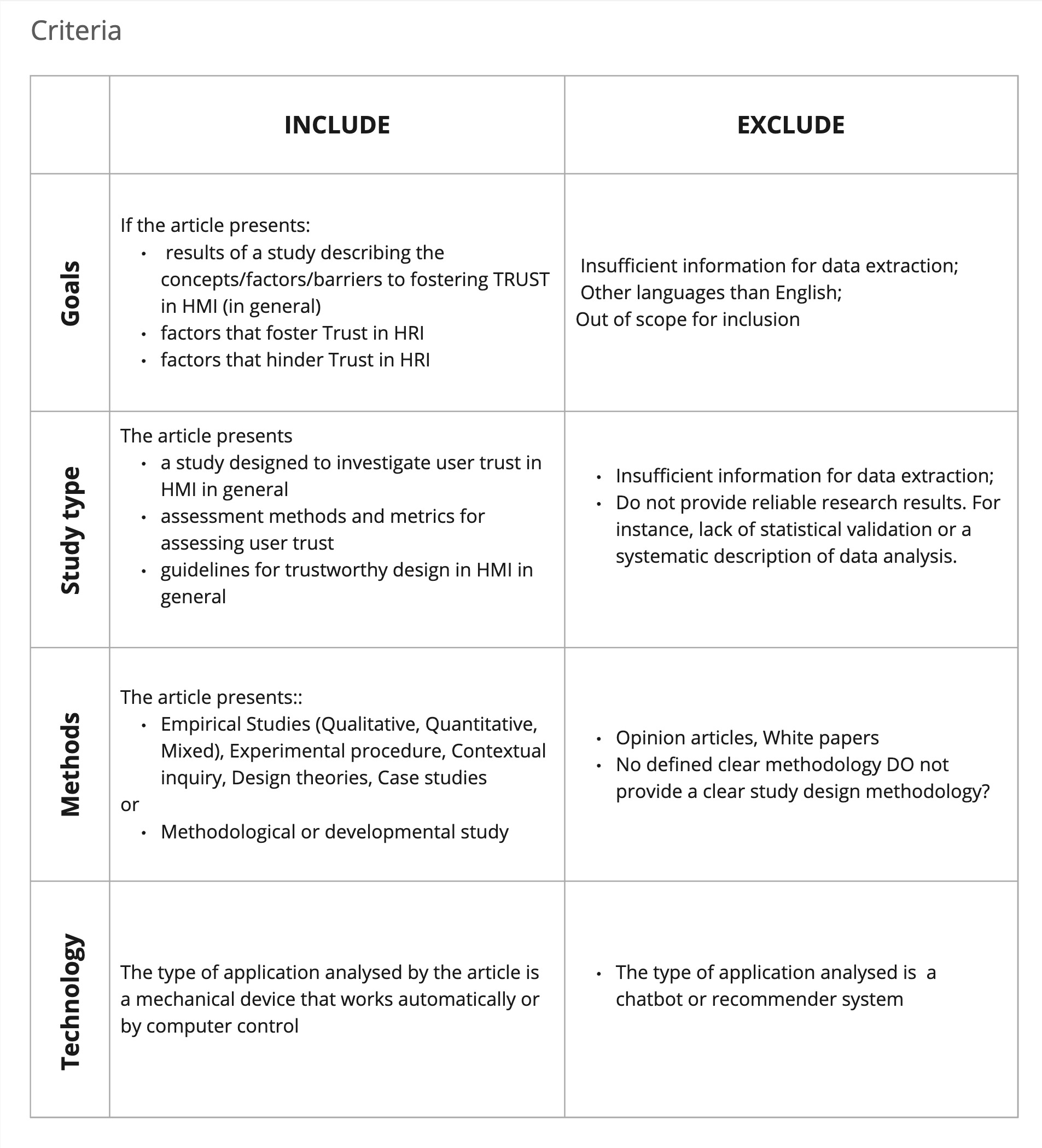}
\caption{The matrix illustrates the inclusion and exclusion criteria applied.\label{fig2}}
\end{figure}   
\unskip

\subsection{Analysis and synthesis}

The dataset from the reviewed articles was organized into different categories to facilitate its analysis, comprehending the employed definitions of trust, the origins and focus of the studies, the kind of application examined, target users, and the methodology deployed. The first and second authors analyzed the aspects listed as facilitators and barriers to trust, using the K.J. method \cite{abeywickrama2023specifying} for clustering and organizing the extracted knowledge.


\section{Analysis \& Results}

The studies summarized in the upcoming sections include diverse geographical regions, applications, and user populations, reflecting the broad scope of trust research in HRI. 

Most reviewed articles have been published in Western Europe (N=12) countries such as Germany, Austria, the UK, and the Netherlands, accounting for over one-third of HRI-trust studies (see Table 1 below). This is followed by Northern America (N=7), mainly the USA, and Eastern Asia (N=6), including China and South Korea. 
Other regions are represented by fewer studies, including Northern Europe (N=2), with studies from Sweden and Norway. Australasia (N=2), primarily Australia, and Southern Europe (N=2), including Spain and Portugal. South-Eastern Asia (Singapore), Eastern Europe (Hungary), and Western Asia (Israel) each contributed one study. We see a lack of studies from several geographical regions, which illustrates the biased nature of research in this field.

\subsection{What are the most common methodologies to study users’ trust in HRI?}

We assessed the types of studies carried out in the HRI field focused on users' trust. More than 59.4\% were experimental studies (N=19). This was followed by systematic literature reviews (N=6), which account for 18.8\% of the reviewed articles, and case studies (N=4), conducted approximately 12.5\% of the time, as illustrated in Figure~\ref{fig3}  below. Delving further into that disproportion, we examined how the HRI studies have defined and assessed users’ trust and in what overarching domains these studies are developed. 

\begin{figure}[!ht]
\includegraphics[width = 11.5cm ]{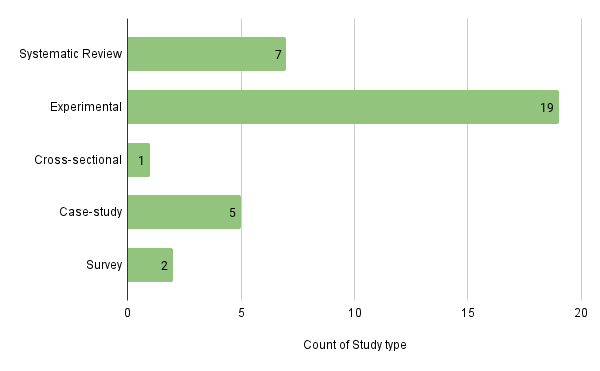}
\caption{Overview of methodologies to study users’ trust in HRI.\label{fig3}}
\end{figure}   
\unskip

\subsubsection{Conceptualization of user trust in HRI}
\begin{figure}[!ht]
\begin{center}
\includegraphics[width = 11cm ]{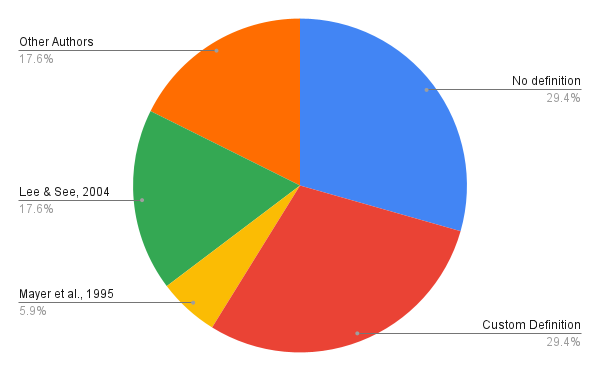}
\caption{Overview of trust definitions adopted by the selected articles.\label{fig4}}
\end{center}
\end{figure}

Figure~\ref{fig4} illustrates the distribution of trust definitions in recent HRI studies. Ten (29.4\%) of the revised articles do not define trust. Likewise, approximately 29.4\% of the reviewed articles apply custom definitions – ad-hoc definitions described by authors either following dictionary expressions or coupling widely used definitions of trust in automation (e.g., widely used definition of trust in automation).

Even serving as a foundational reference for many custom definitions found in our analysis, \cite{lee2004trust} definition of trust is present in only six (17.6\%) of the articles, while \cite{mayer1995trust} Mayer et al. (1995) - even though recognized as a widely used trust definition in the field of HCI – is explicitly adopted in only two (5.4\%) of articles.

\subsection{How do the studies assess trust?}
Considering the methods and strategies deployed to assess trust, we noticed that approximately a third of the reviewed articles (26.5\%) had used Custom Questions (e.g., “Overall, how much would you trust an automated system?” and “How trustworthy did the robot appear to you?”) pertinent to the application investigated by them. These groups also correspond to half of the experimental studies reviewed, as illustrated in Table \ref{tab1}.


\begin{threeparttable}[!ht]
\caption{How trust assessed in reviewed HRI studies}\label{tab1}
\begin{tabular}{p{2.8cm}ccccccccc}\toprule
{\textbf{How the study measures trust?}} &\multicolumn{5}{c}{\textbf{Study Type}} &\multicolumn{2}{c}{\textbf{}} \\\cmidrule{2-8}
&\textbf{Experimental} &\textbf{Syst. Review} &\textbf{Case-study} &\textbf{Survey} &\textbf{Cross-sectional} 
&Total &\% \\\midrule
\textbf{Custom Question} & & & & & &\textbf{} & \\
Daniel et al. (2013) &x & & & & &\textbf{} & \\
Jung et al. (2021) &x & & & & &\textbf{} & \\
Zhu et al. (2023) &x & & & & &\textbf{} & \\
Esterwood et al. (2023) &x & & & & &\textbf{} & \\
Alam et al. (2021) &x & & & & &\textbf{} & \\
Brule et al. (2014) &x & & & & &\textbf{} & \\
Kraus et al. (2023) &x & & & & &\textbf{} & \\
Gaudiello et al. (2016) &x & & & & &\textbf{} & \\
Soh et al. (2018) &x & & & & &\textbf{} & \\
\textbf{Total} & & & & & &\textbf{9} &26.5 \\
\textbf{Combined measures} & & & & & &\textbf{} & \\
Lim et al. (2023) & & &x & & &\textbf{} & \\
Yun et al. (2022) &x & & & & &\textbf{} & \\
Babel et al.(2022) & & &x & & &\textbf{} & \\
Wang et al.(2024) &x & & & & &\textbf{} & \\
Chen et al. (2022) &x & & & & &\textbf{} & \\
Clement et al. (2022) &x & & & & &\textbf{} & \\
Sanders et al. (2018) & & & &x & &\textbf{} & \\
Miller et al. (2021) &x & & & & &\textbf{} & \\
\textbf{Total} & & & & & &\textbf{8} &23.5 \\
\textbf{Validaded Scale} & & & & & &\textbf{} & \\
Pinto et al. (2022) & & & & &x &\textbf{} & \\
Huang et al. (2021) &x & & & & &\textbf{} & \\
Gulati et al. (2019) &x & & & & &\textbf{} & \\
Adami et al. (2022) &x & & & & &\textbf{} & \\
Ambsdorf et al. (2022) & & & &x & &\textbf{} & \\
Kraus et al. (2022) &x & & & & &\textbf{} & \\
Pompe et al. (2022) &x & & & & &\textbf{} & \\
\textbf{Total} & & & & & &\textbf{7} &20.6 \\
\textbf{Self-report} & & & & & &\textbf{} & \\
Cameron et al. (2021) & & &x & & &\textbf{} & \\
Schaefer et al. (2017) & & &x & & &\textbf{} & \\
Koren et al. (2022) & & &x & & &\textbf{} & \\
\textbf{Total} & & & & & &\textbf{3} &8.8 \\
\textbf{N/A} & & & & & &\textbf{} & \\
Alonso et al. (2018) & &x & & & &\textbf{} & \\
Yuan et al. (2021) & &x & & & &\textbf{} & \\
Bach et al. (2024) & &x & & & &\textbf{} & \\
Xu et al. (2023) & &x & & & &\textbf{} & \\
Tian et al. (2021) & &x & & & &\textbf{} & \\
Akalin et al. (2023) & &x & & & &\textbf{} & \\
Schoeller et al. (2021) & &x & & & &\textbf{} & \\

\textbf{Total} & & & & & &\textbf{7} &20.6 \\
\midrule
\textbf{Grand Total} & & & & & &\textbf{34} &100 \\
\bottomrule
\end{tabular}

\end{threeparttable} 


\unskip
In 23.5\% of the reviewed papers, combined measures were used to assess users’ trust in the robots across Experimental, Survey, and Case-study contexts - exploiting pre-and post-testing questionnaires alongside behavioral measurements and self-reporting methods. The study by \cite{Lim_Robb_Wilson_Hastie_2023} mentioned combining questionnaires with user interviews. Also, 20.6\% of the studies used validated tools for empirically measuring user trust across Experimental settings, Surveys, and Cross-sectional studies. Among the assessment instruments, scales such as the Interpersonal Trust Questionnaire (ITQ) by Forbes and Roger (1999) \cite{forbes1999stress}, the Human-Computer Trust Scale (HCTS) by \cite{gulati2019hctm}, and the Trust Perception Scale-HRI \cite{schaefer2013perception} were mentioned. Furthermore, in 8.8\% of the reviewed articles, self-reports (e.g., interviews) were the primary methods used to evaluate users’ trust perceptions of HRI.
This classification did not include the literature reviews and meta-analysis articles by \cite{Alonso_Puente_2018, Yuan_Klavon_Liu_Lopez_Zhao_2021, bach2022reviewtrustai,Xu_Chen_You_2023, Tian_Oviatt_2021, Schoeller_Miller_Salomon_Friston_2021, Akalin_Kiselev_Kristoffersson_Loutfi_2023} - classified below as Non-Applicable.

\subsubsection{Robot deployment domains}
\begin{figure}[!ht]
\begin{center}
   \includegraphics[width = 11.5cm]{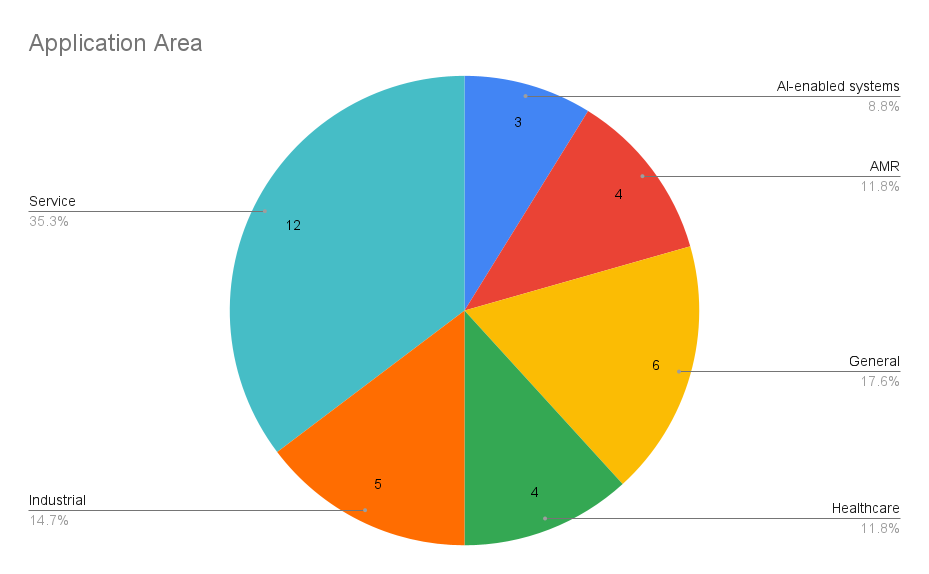}
\caption{The chart illustrates the variety of deployment contexts explored by the selected studies\label{fig6}}

\end{center}
\end{figure}

We analyzed the areas of applications examined by the selected articles. We observed that Service robots - robots that assist humans in professional or personal settings through application areas like social robots, entertainment, and domestic robots - are investigated in ten (35.3\%) of the reviewed articles, as illustrated in Figure \ref{fig6}. 
Eight studies (17.6 \%) explored general aspects of HRI, providing broader perspectives that do not focus on a specific domain. Following, the context of industrial automation was examined in five (14.7\%) of the studies, ranging from manufacturing to construction applications. Autonomous mobile robots (AMR) were investigated in four (11.8\%) of the articles reviewed, exploring both aerial and terrestrial vehicles. Equally, four articles (11.8\%) explored the domain of healthcare with robots implemented for rehabilitation purposes. Additionally, 8.8\% delved into the emergent domain of AI-enabled systems.

\subsection{What has been the focus of HRI researchers when investigating trust?}
While classifying the study focus of the recent HRI studies in human-robot trust, we examined emerging research themes alongside the overarching domains investigated. The research primarily explores human-robot communication and trust-influencing factors, and user studies often emphasize user preferences, safety, and trustworthiness. 
As shown in Table \ref{tab1} below, 23.5\% of the reviewed studies focused on trust indicators and measurements. Some studies, such as those by \cite{gulati2019hctm, bach2022reviewtrustai}, provide broader perspectives by addressing trust across the emergent AI-enabled systems, while others focus on specific applications, such as industrial robots in collaborative application areas examined by \cite{Pinto2022}. In contrast, \cite{Schoeller_Miller_Salomon_Friston_2021} did not specify a domain or area of application but rather provided general perspectives on the theme of extended control.

\begin{threeparttable}[!ht]
\centering
\caption{What have been the focus and application areas of the reviewed HRI studies}\label{tab2}
\scriptsize
\begin{tabular}{p{4.5cm}cccccp{2.5cm}|ccc}\toprule
&\multicolumn{6}{c}{\textbf{Application Area}} & &\textbf{} \\\cmidrule{2-8}
&\textbf{General} &\textbf{Industrial} &\textbf{Service} &\textbf{Healthcare} &\textbf{AMR} &\textbf{AI-enabled systems} &Total &\textbf{\%} \\\midrule
\textbf{Study Focus} &\textbf{} &\textbf{} &\textbf{} &\textbf{} &\textbf{} &\textbf{} & &\textbf{} \\
\midrule
\textbf{Trust indicators and measurements} &\textbf{} &\textbf{} &\textbf{} &\textbf{} &\textbf{} &\textbf{} &\textbf{} &\textbf{} \\
Pinto et al. (2022) & &x & & & & &\textbf{} & \\
Jung et al. (2021) &x & & & & & &\textbf{} & \\
Yun et al. (2022) & & & & &x & &\textbf{} & \\
Bach et al. (2024) & & & & & &x &\textbf{} & \\
Gulati et al. (2019) & & & & & &x &\textbf{} & \\
Koren et al. (2022) & & & &x & & &\textbf{} & \\
Schoeller et al. (2021) &x & & & & & &\textbf{} & \\
Soh et al. (2018) & & &x & & & &\textbf{} & \\
\textbf{Total} & & & & & & &\textbf{8} &23.5 \\
\textbf{Human-robot communication} &\textbf{} &\textbf{} &\textbf{} &\textbf{} &\textbf{} &\textbf{} &\textbf{} &\textbf{} \\
Alonso et al. (2018) &x & & & & & &\textbf{} & \\
Lim et al. (2023) & & &x & & & &\textbf{} & \\
Huang et al. (2021) & & &x & & & &\textbf{} & \\
Tian et al. (2021) &x & & & & & &\textbf{} & \\
Ambsdorf et al. (2022) & & & & & &x &\textbf{} & \\
Kraus et al. (2022) & & &x & & & &\textbf{} & \\
Pompe et al. (2022) & & &x & & & &\textbf{} & \\
\textbf{Total} & & & & & & &\textbf{7} &20.6 \\
\textbf{User studies} &\textbf{} &\textbf{} &\textbf{} &\textbf{} &\textbf{} &\textbf{} &\textbf{} &\textbf{} \\
Daniel et al. (2013) & &x & & & & &\textbf{} & \\
Yuan et al. (2021) & & & &x & & &\textbf{} & \\
Adami et al. (2022) & &x & & & & &\textbf{} & \\
\textbf{Total} & & & & & & &\textbf{3} &8.8 \\
\textbf{Human-robot teaming} &\textbf{} &\textbf{} &\textbf{} &\textbf{} &\textbf{} &\textbf{} &\textbf{} &\textbf{} \\
Schaefer et al. (2017) & & & & &x & &\textbf{} & \\
Zhu et al. (2023) & & & & &x & &\textbf{} & \\
\textbf{Total} & & & & & & &\textbf{2} &5.9 \\
\textbf{Robot behaviors} &\textbf{} &\textbf{} &\textbf{} &\textbf{} &\textbf{} &\textbf{} &\textbf{} &\textbf{} \\
Brule et al. (2014) & & &x & & & &\textbf{} & \\
\textbf{Total} & & & & & & &\textbf{1} &2.9 \\
\textbf{Human-robot communication \& User studies} &\textbf{} &\textbf{} &\textbf{} &\textbf{} &\textbf{} &\textbf{} &\textbf{} &\textbf{} \\
Babel et al.(2022) & & &x & & & &\textbf{} & \\
Alam et al. (2021) & & & &x & & &\textbf{} & \\
Gaudiello et al. (2016) & & &x & & & &\textbf{} & \\
\textbf{Total} & & & & & & &\textbf{3} &8.8 \\
\textbf{User studies \& Trust indicators and measurements} &\textbf{} &\textbf{} &\textbf{} &\textbf{} &\textbf{} &\textbf{} &\textbf{} &\textbf{} \\
Clement et al. (2022) & & & & &x & &\textbf{} & \\
Cameron et al. (2021) & & &x & & & &\textbf{} & \\
Miller et al. (2021) & & &x & & & &\textbf{} & \\
\textbf{Total} & & & & & & &\textbf{3} &8.8 \\
\textbf{Human-robot communication \& Robot behaviors} &\textbf{} &\textbf{} &\textbf{} &\textbf{} &\textbf{} &\textbf{} &\textbf{} &\textbf{} \\
Esterwood et al. (2023) & &x & & & & &\textbf{} & \\
Kraus et al. (2023) & & &x & & & &\textbf{} & \\
\textbf{Total} & & & & & & &\textbf{2} &5.9 \\
\textbf{Human-robot communication \& Trust indicators and measurements} &\textbf{} &\textbf{} &\textbf{} &\textbf{} &\textbf{} &\textbf{} &\textbf{} &\textbf{} \\
Xu et al. (2023) &x & & & & & &\textbf{} & \\
\textbf{Total} & & & & & & &\textbf{1} &2.9 \\
\textbf{User studies \& Human-robot teaming} &\textbf{} &\textbf{} &\textbf{} &\textbf{} &\textbf{} &\textbf{} &\textbf{} &\textbf{} \\
Sanders et al. (2018) & &x & & & & &\textbf{} & \\
\textbf{Total} & & & & & & &\textbf{1} &2.9 \\
\textbf{Robot behaviors \& Trust indicators and measurements} &\textbf{} &\textbf{} &\textbf{} &\textbf{} &\textbf{} &\textbf{} &\textbf{} &\textbf{} \\
Chen et al. (2022) & & & &x & & &\textbf{} & \\
\textbf{Total} & & & & & & &\textbf{1} &2.9 \\
\textbf{Robot behaviors \& Perceived safety} &\textbf{} &\textbf{} &\textbf{} &\textbf{} &\textbf{} &\textbf{} &\textbf{} &\textbf{} \\
Wang et al.(2024) & & &x & & & &\textbf{} & \\
\textbf{Total} & & & & & & &\textbf{1} &2.9 \\
\textbf{Human-robot teaming \& Perceived safety} &\textbf{} &\textbf{} &\textbf{} &\textbf{} &\textbf{} &\textbf{} &\textbf{} &\textbf{} \\
Akalin et al. (2023) &x & & & & & &\textbf{} & \\
\textbf{Total} & & & & & & &\textbf{1} &2.9 \\
\midrule
\textbf{Grand Total} & & & & & & &\textbf{34} &100.0 \\
\bottomrule
\end{tabular}
\end{threeparttable}

\unskip

Another focus of interest explored was human-robot communication, representing 20.6\% of the results. \cite{Alonso_Puente_2018} offered a meta-analysis on transparency in the framework of shared autonomy. And \cite{Tian_Oviatt_2021} investigated robot ’errors as marks of socio-affective competence. Moreover, \cite{Lim_Robb_Wilson_Hastie_2023, Huang_Rau_Ma_2021, Kraus_Wagner_Untereiner_Minker_2022, Pompe_Velner_Truong_2022} focused on human-robot communication in the service robotics domain observing applications such as robot barista \cite{Lim_Robb_Wilson_Hastie_2023}  and catering robot \cite{Kraus_Wagner_Untereiner_Minker_2022}. 
Moreover, 8.8\% of the HRI articles focused on user experiences and characteristics. In investigating industrial autonomation, \cite{Daniel_Thomessen_Korondi_2013} observed user preferences for industrial robots' interfaces, and \cite{Adami_Rodrigues_Woods_Becerik-Gerber_Soibelman_Copur-Gencturk_Lucas_2022} focused on the application of construction robots. 
While \cite{Zhu_Wang_Wang_Quan_Tang_2023} and \cite{Schaefer_Straub_Chen_Putney_Evans_2017} focused on human-robot teaming in the context of Autonomous mobile robots (AMR). \cite{Brule_Dotsch_Bijlstra_Wigboldus_Haselager_2014} explored the implications of robot behavior styles in the context of service robots for domestic applications. \cite{Xu_Chen_You_2023} explored human-robot communication and trust indicators and measurements while examining social cues for general HRI applications. Moreover, \cite{Sanders_Kaplan_Koch_Schwartz_Hancock_2018} focused on User studies \& Human-robot teaming, analyzing users’ reliance on robot outputs. 
Perceived safety was a focus area of two of the reviewed studies. In \cite{Wang_Wang_Ge_Duan_Chen_Wen_2024} recent study, they explored Robot behaviors \& Perceived safety in the service domain, investigating soft robots for social and entertainment applications. The study by\cite{Akalin_Kiselev_Kristoffersson_Loutfi_2023} proposed a taxonomy focused on Human-robot teaming \& Perceived safety.

\subsection{What are the barriers and facilitators for fostering trustworthy HRI?}

\begin{figure}[!ht]
\includegraphics[width=16.48cm]{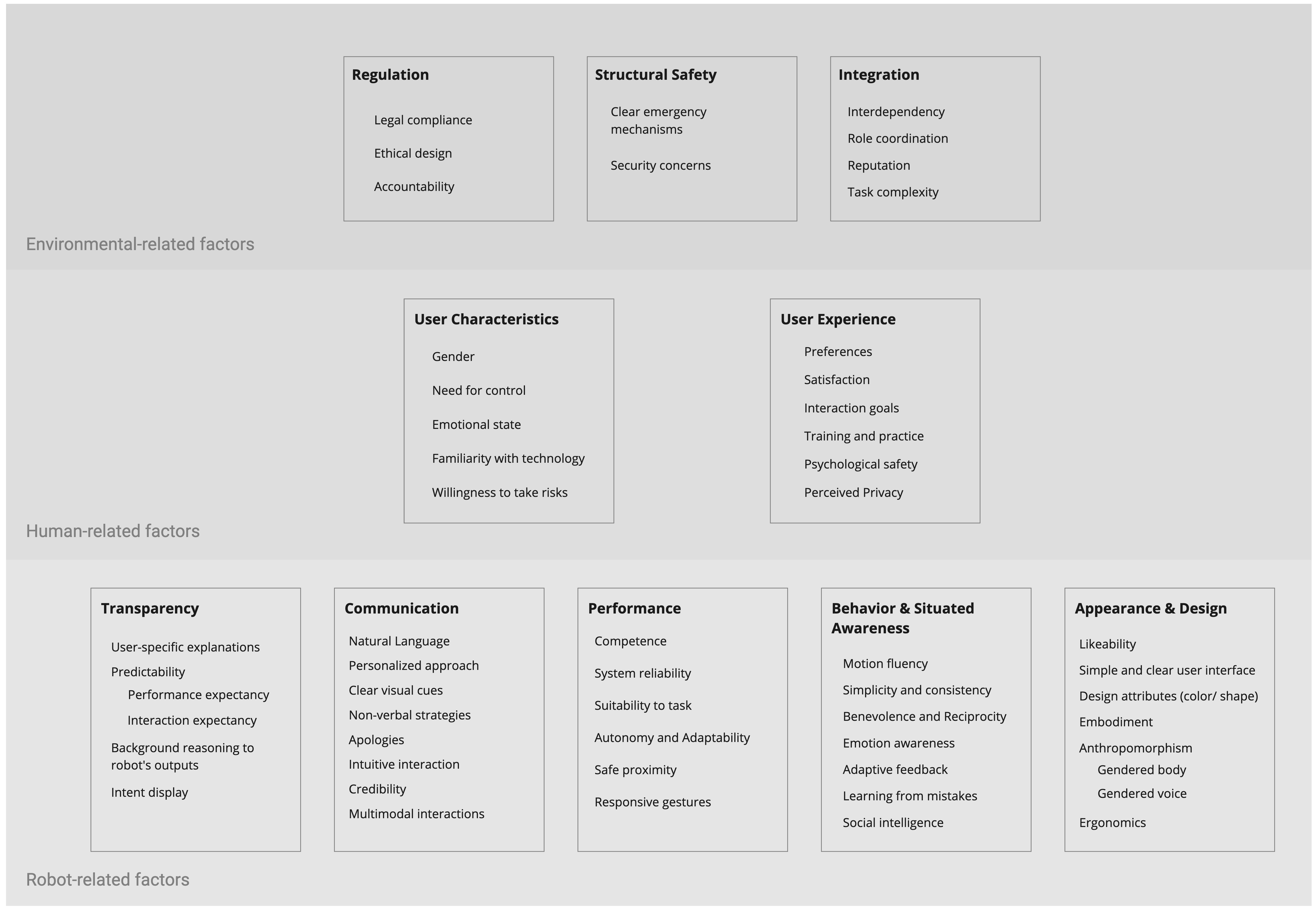}
\caption{The diagram shows the elements of trust facilitators identified by the reviewers\label{fig7}}
\end{figure}

\begin{figure}[!ht]
\includegraphics[width=16.48cm]{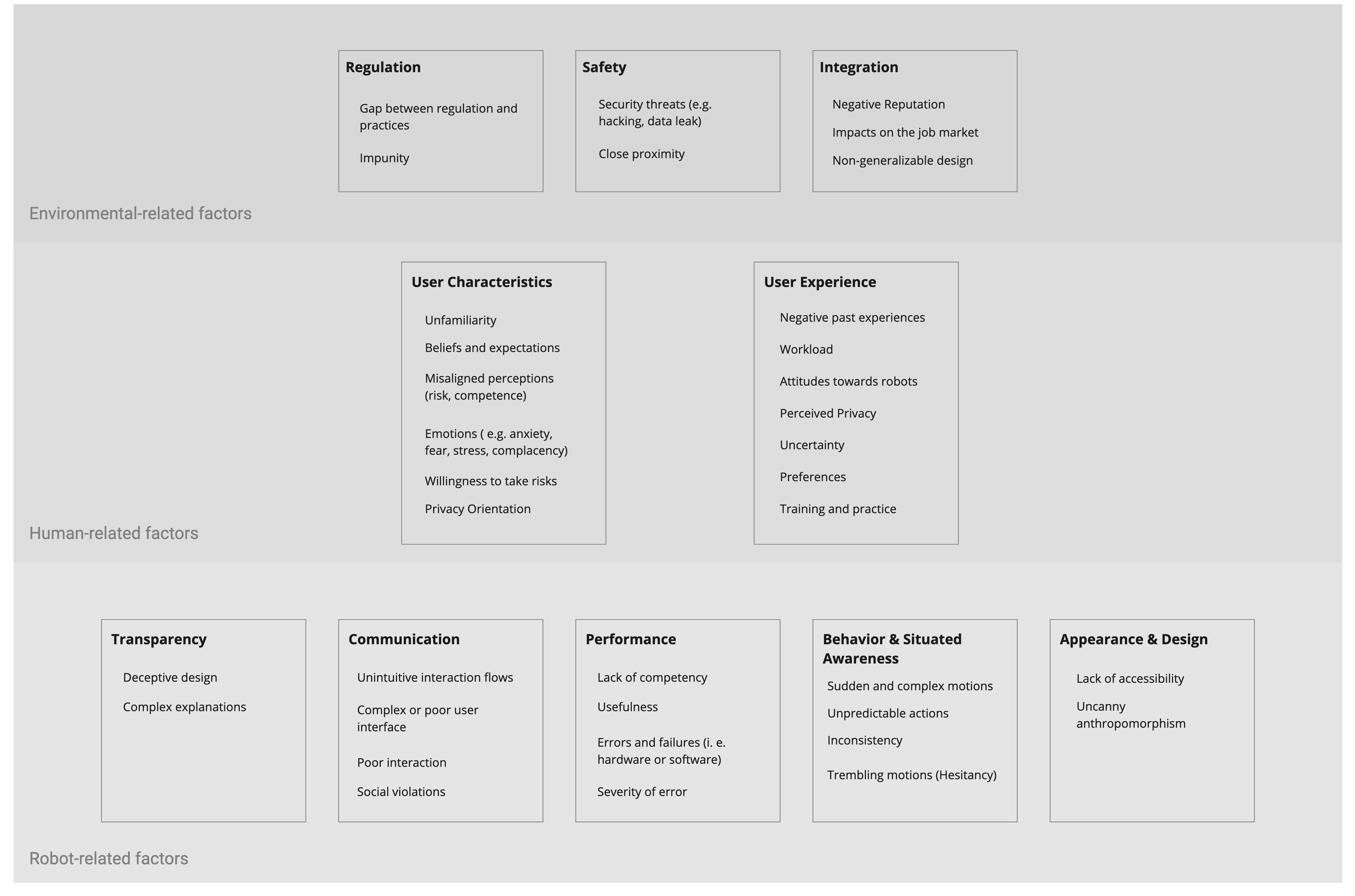}
\caption{The diagram shows the elements of trust facilitators identified by the reviewers\label{fig8}}
\end{figure}

We adopted Hancock et al.’s \cite{hancock2011can} classification to explore the factors influencing users' perceptions of trust in Human-Robot Interaction (HRI). In this classification, Hancock and colleagues identify trust-predicting factors related to human, robot, and environmental characteristics within HRI. \cite{Schaefer_Straub_Chen_Putney_Evans_2017} noted that due to trust's dynamic nature, it is essential to understand the variety of potential design cues and functionalities that affect trust and, consequently, support the integration of autonomous robots in diverse scenarios. With that in mind and intrigued by possible mechanisms to calibrate appropriate levels of trust during interactions, we proposed a breakdown of the factors to clarify potential constraints (i.e., barriers) and enablers (i.e., facilitators) related to user needs, design, functionalities, and contextual elements that can be utilized to mediate trust in HRI. Presented in the diagrams below (Figures \ref{fig7} and \ref{fig8}) are the key attributes, processes, and contextual considerations relevant to the design and operation of trustworthy HRI systems. Each category was developed from the topics emerging from our data extraction and organized by the first and second authors into trust facilitators and barriers. It follows the rationale of:

\begin{enumerate}
\item Robot-related factors as the attributes of robotic systems divided into Transparency, Communication, Performance, Behavior \& Situated Awareness, and Appearance \& Design; 
\item Human-related factors, which refer to user characteristics and their experiences; and
\item Environmental-related factors representing contextual aspects that include Regulation, Safety, and Integration considerations.
\end{enumerate}

Our analysis observed that factors influencing trust in automation can act as constraints and enablers. These factors are deeply interwoven with the context of the application, and they may vary depending on the specifics of the user, type of task, and type of robot. This highlights the inherent complexity in studying trust as a quality of the user experience within interaction with autonomous robots. 
For instance, anthropomorphism, one aspect of robot-related factors, is considered a facilitator in service and healthcare domains where social interaction occurs and adaptive feedback mechanisms are expected \cite{Yuan_Klavon_Liu_Lopez_Zhao_2021, Lim_Robb_Wilson_Hastie_2023}. 
As well as human-like features, such as the robot's use of natural language \cite{Koren_Polak_Levy-Tzedek_2022}, and socio-affective competence and awareness of social rules \cite{Cameron_Saille_Collins_Aitken_Cheung_Chua_Loh_Law_2021} – often portraying application areas of social robots for companionship and rehabilitation \cite{Kraus_Wagner_Untereiner_Minker_2022, Pompe_Velner_Truong_2022, Alam_Johnston_Vitale_Williams_2021} -  are referred as influencing the user perception of the robot's intelligence \cite{Tian_Oviatt_2021, Chen_Zhai_Liu_2022}. Moreover, \cite{Ambsdorf} point out that gendering a robot’s voice can affect the robot's perception in terms of stereotypes, preferences, and trust.

\unskip

On the other hand, studies investigating applications in the industrial and transportation sector shed light on facilitating trust through factors such as simple and clear user interface \cite{Daniel_Thomessen_Korondi_2013}, intent display \cite{Yun_Yang_2022}, and suitability to task \cite{clement2022enhancing}. Although complementary perspectives on implementing robots in these contexts have also highlighted aspects of social interaction, such as emotion awareness \cite{Esterwood_Robert_2023} and benevolence \cite{Pinto2022}.

Aspects of a robot's performance and behavior are considered to facilitate trust across different domains of robotics. Besides system reliability \cite{Soh_Xie_Chen_Hsu_2018, bach2022reviewtrustai}, the robot's motion fluency \cite{Babel_Kraus_Baumann_2022, Brule_Dotsch_Bijlstra_Wigboldus_Haselager_2014} and the simplicity and consistency in robot's behavior \cite{Pinto2022, Wang_Wang_Ge_Duan_Chen_Wen_2024} can influence users trust. Simplified motions have been shown to alleviate user anxiety, particularly among individuals with limited technology acceptance \cite{Wang_Wang_Ge_Duan_Chen_Wen_2024}. Moreover, the robot's ability to learn from mistakes (recognizing and recovering from them) might impact user engagement and interaction, potentially even turning this into a net positive for the interaction \cite{Cameron_Saille_Collins_Aitken_Cheung_Chua_Loh_Law_2021}. Considering the human-related factors, \cite{Akalin_Kiselev_Kristoffersson_Loutfi_2023} refer to psychological safety as an aspect of the user experience that might enable trust in human-robot teaming. \cite{Lim_Robb_Wilson_Hastie_2023} and \cite{Miller_Kraus_Babel_Baumann_2021} pointed out the benefit of tailoring interactions to the user preferences and needs in the service robot's domain. Similarly, \cite{Yun_Yang_2022} identifies tailored interaction as a trusted facilitator in the context of autonomous mobile robots (AMR). 

\unskip

User characteristics such as familiarity with technology are noted by \cite{Wang_Wang_Ge_Duan_Chen_Wen_2024, Gaudiello_Zibetti_Lefort_Chetouani_Ivaldi_2016, Kraus_Merger2023} as a factor in enabling human trust in the robot over varied applications of service robots. Likewise, \cite{Adami_Rodrigues_Woods_Becerik-Gerber_Soibelman_Copur-Gencturk_Lucas_2022} noted that it could enhance situational awareness and mental load during task execution alongside construction robots.
Furthermore, authors indicate gender \cite{clement2022enhancing, Ambsdorf, Koren_Polak_Levy-Tzedek_2022, Schaefer_Straub_Chen_Putney_Evans_2017}, emotional state \cite{Alam_Johnston_Vitale_Williams_2021, Kraus_Merger2023}, need for control \cite{Wang_Wang_Ge_Duan_Chen_Wen_2024, Schoeller_Miller_Salomon_Friston_2021, Gaudiello_Zibetti_Lefort_Chetouani_Ivaldi_2016} and willingness to take risks \cite{Pinto2022, clement2022enhancing} as trust influencing factors in HRI. 

Among the environmental-related factors, authors mention that trust can be facilitated by regulatory means such as legal compliance \cite{bach2022reviewtrustai, Esterwood_Robert_2023} and ethical design \cite{Alonso_Puente_2018, Yun_Yang_2022}. As well as by ensuring structural safety with precise emergency mechanisms \cite{Wang_Wang_Ge_Duan_Chen_Wen_2024}.\cite{Adami_Rodrigues_Woods_Becerik-Gerber_Soibelman_Copur-Gencturk_Lucas_2022} indicated interdependency as a trusted facilitator for integrating industrial robots in construction applications, which is also pointed out as a trusted facilitator for integrating military systems \cite{Schaefer_Straub_Chen_Putney_Evans_2017}. Task complexity is noted across application areas such as autonomous aerial vehicles \cite{Zhu_Wang_Wang_Quan_Tang_2023}, COBOTS \cite{Sanders_Kaplan_Koch_Schwartz_Hancock_2018}, and domestic robots \cite{Soh_Xie_Chen_Hsu_2018}.

A close analysis of the barriers identified in the reviewed studies revealed that among the environment-related factors, \cite{bach2022reviewtrustai} emphasized the gap between regulations and practice as an aspect that hinders user trust in AI-enabled systems. Similarly, \cite{Alonso_Puente_2018} identified impunity as a barrier to trust and proposed equipping robots with ethical black box mechanisms to address this issue to enhance accountability. Furthermore, the authors motioned negative reputation as a trust barrier in autonomous technologies. Inserted in a socio-technical system, robots are constantly subject to society's evaluation. Taking the context of AMR, \cite{Yun_Yang_2022} emphasized that adverse reactions toward autonomous vehicles might be fueled by the media highlights regarding accidents involving these systems.

Considering Human-related factors, the reviewed articles mentioned trust barriers such as misaligned perceptions \cite{Pinto2022} and user emotions like anxiety \cite{Wang_Wang_Ge_Duan_Chen_Wen_2024, Miller_Kraus_Babel_Baumann_2021}, fear and stress \cite{clement2022enhancing}, and complacency \cite{Schoeller_Miller_Salomon_Friston_2021}. Characteristics of user privacy orientation and perceived privacy are emphasized by the works of \cite{gulati2019hctm} and \cite{Ambsdorf}. 

Reflecting on robot-related factors, we noticed that studies tackling service and industrial domains indicated that in application areas demanding human-robot collaboration, lack of competency \cite{Brule_Dotsch_Bijlstra_Wigboldus_Haselager_2014, Cameron_Saille_Collins_Aitken_Cheung_Chua_Loh_Law_2021}, hesitant behavior (i.e., trembling motions) \cite{Chen_Zhai_Liu_2022}, and failed hardware or software performance \cite{Kraus_Wagner_Untereiner_Minker_2022, Soh_Xie_Chen_Hsu_2018, Akalin_Kiselev_Kristoffersson_Loutfi_2023} negatively impact users’ trust. Similarly, \cite{Tian_Oviatt_2021} highlighted that social norm violations harm the user’s perception of a robot’s socio-affective competence. Moreover, \cite{Koren_Polak_Levy-Tzedek_2022} remarked that the severity of errors may also impact trust perception to a different degree.

\section{Conclusions}

This literature review was intended to provide insights into how human-robot trust is conceptualized and approached within recent HRI studies. The conclusions described here were crafted to aid designers and developers in identifying and reflecting on critical barriers and facilitators to trust in HRI. 
Our findings illustrate a shortfall in using accredited trust definitions in recent years. Approximately a third of the reviewed studies touch upon trust-related subject matters but do not explicitly define the concept. Trust is a phenomenon investigated across diverse disciplines, which, to our knowledge, brings such a tendency to use competing and often contradictory definitions, models, and frameworks. 
\cite{bach2022reviewtrustai} proposed that selecting the most appropriate trust definition concerning the context investigated is better than pursuing a unified trust definition or comparing the existing ones.  Nonetheless, our analysis demonstrates that almost 30\% of the revised articles do not present a definition of trust, which can be problematic as trust is a phenomenon interwoven and affected by other interaction factors. Moreover, our findings endorse that the conceptual gap in trust in technology literature persists and that this lack of conceptual clarity explains the sometimes-contradictory findings in the literature \cite{gulati2019hctm, Schaefer_Straub_Chen_Putney_Evans_2017, sousa2023challenges, hancock2011meta, hoff2015trust}.
As shown by our analysis, this might also reflect the choice of methods and instruments used to investigate trust in robots and other autonomous systems. The results demonstrated that most studies assess trust by custom questions – often a single likelihood assessment – where users rate the overall experience but do not reflect on the related constructs that impact their perceptions of trust. Although that can be argued as a strategy to overcome disruptions in the course of interaction, 23\% of the studies demonstrated that combined measures can be deployed for a comprehensive assessment. \cite{Sanders_Kaplan_Koch_Schwartz_Hancock_2018} combined a wide range of methods, including demographics and personality questions, alongside the Negative Attitudes Toward Robots Scale (NARS) \cite{nomura2006measurement}, the Interpersonal Trust Questionnaire (ITQ) \cite{forbes1999stress}, and the Trust in Automated Systems Survey \cite{jian2000foundations}. Pompe, et al. \cite{Pompe_Velner_Truong_2022} also collected measures across different environments and contexts using validated scales, including the Godspeed Questionnaire \cite{bartneck2023godspeed}, to understand the factors affecting user perception of robots.

Recent studies by \cite{Alam_Johnston_Vitale_Williams_2021, Huang_Rau_Ma_2021, Cameron_Saille_Collins_Aitken_Cheung_Chua_Loh_Law_2021, Esterwood_Robert_2023} have emphasized the importance of human-centric drive for autonomous technology development, underscoring the value of designing systems to communicate trustworthiness through optimized usability and increased transparency. Emerging themes include perceived safety and human-robot teaming in collaborative and service robots, with increased efforts to balance technical innovation with human-centric design principles for better observability and predictability of system behavior, as shown by \cite{Wang_Wang_Ge_Duan_Chen_Wen_2024} and \cite{Alonso_Puente_2018}.

As described in our analysis, user characteristics and perceptions of the interaction – including individual preferences and needs, how they take risks, psychological safety, control, emotional states, and familiarity with robots - are pivotal as trust facilitators and barriers in HRI. Hence, tailoring interactions to individual needs might influence trust and affect engagement and adoption rates, as commented by \cite{Miller_Kraus_Babel_Baumann_2021, Kraus_Merger2023}. Providing tailored information and explanations is also an essential aspect, as seen in the studies by \cite{Babel_Kraus_Baumann_2022,Pinto2022, bach2022reviewtrustai}. Furthermore, it was seen that negative experiences or unmet expectations can significantly affect usability and acceptance, emphasizing the need for engaging and intuitive HRI designs \cite{Yun_Yang_2022, Kraus_Wagner_Untereiner_Minker_2022}.

As mechanisms to repair user trust, robots should exhibit accountable behaviors, offer adaptive feedback, and implement trust-repair strategies. As noted by recent studies with social robots for service and healthcare, physical appearance and human-like attributes also enhance trust and acceptance \cite{Cameron_Saille_Collins_Aitken_Cheung_Chua_Loh_Law_2021, Yuan_Klavon_Liu_Lopez_Zhao_2021, Jung_Cho_Kim_2021}. Furthermore, the broader context of interactions, including shared control and ethical considerations, influences user trust. Design must account for socio-ethical concerns, privacy, and security in shared environments \cite{Schaefer_Straub_Chen_Putney_Evans_2017,Alonso_Puente_2018,Ambsdorf}. 
On the other hand, robot errors, performance failures, and lack of competency might negatively affect trust and reliability in robot actions \cite{Tian_Oviatt_2021,Cameron_Saille_Collins_Aitken_Cheung_Chua_Loh_Law_2021, Kraus_Merger2023}. Additionally, the studies by \cite{Yuan_Klavon_Liu_Lopez_Zhao_2021}, \cite{Schoeller_Miller_Salomon_Friston_2021}, and \cite{Zhu_Wang_Wang_Quan_Tang_2023} also indicate system design issues as barriers to trust in HRI, demonstrating that complex explanations, poor interaction flows, and unpredictable behavior contribute to user frustration and hinder user trust.

Our literature review highlights that the domain and area of application may impact how humans perceive robot trustworthiness. Distinct factors may prove more influential in specific use scenarios, suggesting a tailored approach to fostering trust in HRI is necessary. It is crucial to delineate when trust pertains to interaction attributes—such as task appropriateness and machine suitability—or when it reflects user preferences, which, as our findings suggest, significantly influence their perceptions of the information exchange and their assessment of robot capabilities.

Addressing technical and human-centered aspects while integrating AI and robotic systems into human environments has become increasingly demanding. This includes ensuring system security, efficiency, flexibility, and resilience, while also addressing human-related qualities such as trust, psychological safety, and ethical considerations \cite{IBM, Shneiderman:2020hv, nahavandi2019industry, Schoeller_Miller_Salomon_Friston_2021}. The foundational components of the European Commission's framework for this symbiotic collaboration contemplate a shift in industrial priorities, ensuring human-centric and sustainable approaches for efficient industrial innovation. For society, this means a perspective for dealing with emerging societal trends and needs and a shift in industrial priorities, emphasizing human-centric and sustainable approaches alongside technological innovation. For research, this represents an opportunity to observe technology emergency and adoption phenomenon from a multidisciplinary approach. 

Researchers, furthermore, highlighted the need for a distinct understanding of trust formation and calibration, calling attention to the different stages and dynamics of trust before and during the interaction with a robot, which is considered “to emerge in an attitude-formation process and then be calibrated along with a comparison of expectations with a robot’s behavior” \cite{Miller_Kraus_Babel_Baumann_2021},p. 03. Prolonged interaction over time can redress issues of lack of trust - \cite{Wang_Wang_Ge_Duan_Chen_Wen_2024, Koren_Polak_Levy-Tzedek_2022, Kraus_Merger2023, Gaudiello_Zibetti_Lefort_Chetouani_Ivaldi_2016} demonstrated the role of familiarity in increasing trust in extended interactions as users become more familiar with the robot’s behaviors and capabilities. 
Our review also contemplated the recommendations for collaboration between humans and robots presented by \cite{Schoeller_Miller_Salomon_Friston_2021}. The authors explain that maximizing trust is not the recommended strategy for successfully integrating robots, as that will also involve addressing the implications of disuse (i.e., missed opportunity) and misuse (i.e., blind trust) of autonomous systems. Therefore, designers and developers must find a balanced level of trustworthiness conducive to effective and sustainable human-robot trust in practical, real-world contexts.

In this literature review, we argue that besides building systems that ensure trustworthiness by attending to requirements of safety and security, it is vital to address the non-functional properties of the interaction concerning the user perceptions of trust and the factors impacting such perception. We observed a divergence in the backgrounds forming the basis for various studies on trust in HRI. An open question remains whether the aspects investigated align consistently with the major factors we identified earlier and whether a synthesis of attributes from different categories could enhance machine adaptability. This synthesis could potentially preempt misuse, accommodate unforeseen changes, and even proactively engage with humans to augment decision-making capabilities and physical interactions per Industry 5.0 principles.

\subsection{Future considerations}

Evolving collaboration between humans and machines requires enabling rapid and often high-stakes decision-making in the industrial, healthcare, government, and military sectors. It is pivotal to ensure that the industrial sector adheres to sustainability practices. Moreover, technology must adjust to align with human values, skills, and well-being, ultimately fostering effective and efficient collaboration between humans and machines.
Providing users with relevant and reliable information about a system's work can increase user trust. This may include explanations of algorithms, model performance, risk factors, contextual information, and actionable information. The amount of information should not be overwhelming and should be targeted to specific user groups. Another critical aspect is the resilience of systems to adapt to unexpected disruptions, such as pandemics or geopolitical challenges.
We aimed to contribute to Industry 5.0 human-centric vision by providing a literature overview of the potential trust-influencing factors that can hinder or facilitate trustworthy human-machine collaborations. Identifying the factors that positively or negatively influence trust in autonomous robots is challenging. Furthermore, quantifying the impact of individual factors, or the cumulative effect of various combinations, proves to be a complex task. Due to the general and abstract nature of our current classification system, it is impractical to directly derive mechanisms for calibrating trust from it. However, this framework serves as a valuable tool for researchers. It provides a foundation for selecting and investigating specific aspects of trust in human-robot interaction (HRI), guiding future studies in this evolving field. These findings can inform future technical and design strategies, research, and initiatives that foster and maintain human trust in autonomous robots. 

\subsection{Limitations}

This systematic literature review explored how trust has been conceptualized and studied as a pathway to facilitate the successful implementation of autonomous technologies. In doing so, we examined the primary definitions adopted by researchers and the methods used to assess trust in HRI systems. Moreover, our overview offers a classification of trust-related barriers and facilitators examined in recent HRI studies. Nonetheless, we must acknowledge limitations in our work, hence reporting on the overcoming strategies taken. 
In this study, we implemented a standardized approach for literature review and data extraction. This approach had the advantage of providing a framework for the reviews; however, it may have constrained the reviewers' appreciation of the data. To remedy this limitation, we held periodical meetings to align the activity’s progress. Another potential issue refers to biases caused by the choices of search keywords, selection criteria, and researchers' interpretations of the selected literature.
As commented in the previous sections, limitations may also result from the article's geographic location and the underrepresentation of certain regions in the poll of selected studies. We try to address these by detailing the article’s affiliation.
Furthermore, the quality of the reviewed studies could potentially limit the content extracted and analyzed in this review. As we undertake a multidisciplinary review of recent HRI studies through the lens of trust in technology, we foresee a potential thematic gap that could impact our research findings.

\paragraph{Authors individual contributions Conceptualization, D.S. and S.S.; methodology, D.S. and S.S.; 
validation, all; formal analysis, D.S. and S.S.; writing---original draft preparation, D.S. and S.S.; writing---review and editing, all; supervision, S.S.; project administration, S.S and O.D; funding acquisition, O.D. All authors have read and agreed to the published version of the manuscript.}

\paragraph{This research was supported by the project \emph{"Increasing the knowledge intensity of Ida-Viru entrepreneurship“} co-funded by the European Union}



\bibliographystyle{apalike} 
\bibliography{template.bib}

\end{document}